\documentclass[12pt,paper]{aastex}

\usepackage{graphicx,natbib,textcomp,caption,subcaption}         
\usepackage[utf8]{inputenc}
\usepackage{xcolor,amsmath}
\usepackage{multirow}
\usepackage{float}
\usepackage{hyperref}
\usepackage{xfrac}

\slugcomment{ApJ manuscript, \today}

\begin{document}

\title{Standing kink waves in sigmoid solar coronal loops: implications for coronal seismology}

\author{N. Magyar, V. M. Nakariakov}
\affil{Centre for Fusion, Space and Astrophysics, Physics Department, University of Warwick, Coventry CV4 7AL, UK; norbert.magyar@warwick.ac.uk}

\begin{abstract}
Using full three-dimensional magnetohydrodynamic numerical simulations, we study the effects of magnetic field sigmoidity or helicity on the properties of the fundamental kink oscillation of solar coronal loops. Our model consists of a single denser coronal loop, embedded in a plasma with dipolar force-free magnetic field with a constant $\alpha$-parameter.
For the loop with no sigmoidity, we find that the numerically determined oscillation period of the fundamental kink mode matches the theoretical period calculated using WKB theory. In contrast, with increasing sigmoidity of the loop, the actual period is increasingly smaller than the one estimated by WKB theory. Translated through coronal seismology, increasing sigmoidity results in magnetic field estimates which are increasingly shifting towards higher values, and even surpassing the average value for the highest $\alpha$ value considered. Nevertheless, the estimated range of the coronal magnetic field value lies within the mimimal/maximal limits, proving the robustness coronal seismology. We propose that the discrepancy in the  estimations of the absolute value of the force-free magnetic field could be exploited seismologically to determine the free energy of coronal loops, if averages of the internal magnetic field and density can be reliably estimated by other methods. 
\end{abstract}


\section{Introduction}

Coronal loops are the building blocks of the closed solar corona, appearing as bright arcs of plasma when viewed in extreme ultraviolet \citep[e.g.,][]{2014LRSP...11....4R}. Understanding their formation and dynamics is essential in solving the long-standing coronal heating problem \citep[e.g.,][]{2012RSPTA.370.3217P}. 
Large-amplitude transverse (kink) oscillations of coronal loops were first reported by \cite{1999ApJ...520..880A} and \cite{1999Sci...285..862N}. These observations allowed for the previously theorized tool of coronal seismology \citep{1970PASJ...22..341U,1984ApJ...279..857R}, applied first by \cite{2001A&A...372L..53N}, to emerge \citep[see,][for comprehensive reviews]{2005RSPTA.363.2743D,2012RSPTA.370.3193D,2012scsd.book.....S}. In coronal seismology, observed wave properties, such as wave period, are compared to theoretical results, allowing plasma parameters which are hard to measure otherwise, such as magnetic field strength, to be inferred. The detailed knowledge of plasma properties is needed e.g. for advancing the prediction of space weather events \citep[e.g.,][]{2010SGeo...31..581S,2017SSRv..212.1253L,2017SSRv..212.1137K}. \par
For the first application of coronal seismology, the simplest theoretical model of a coronal loop was used \citep{1975...37..3,1983SoPh...88..179E}, comprising of a straight flux tube with different constant Alfv\'en speeds inside and outside of the tube. In what followed, several additional effects which might change the oscillation properties were considered, such as varying Alfv\'en speed along the loop \citep{2005SoPh..229...79D,2005A&A...430.1109A,2005A&A...441..361A}, variable cross-section \citep{2008ApJ...686..694R,2008A&A...486.1015V}, elliptical cross-section \citep{2003A&A...409..287R}, loop curvature \citep{2004A&A...424.1065V,2009SSRv..149..299V}, and twisted magnetic field \citep{2007SoPh..246..119R,2012A&A...548A.112T,2015A&A...580A..57R}, among others. \par
Oscillations in coronal loops which exhibited non-planar geometry were first observed by \citet{2002SoPh..206...69S}. Later it was determined that a significant portion of coronal loops are non-planar, i.e. exhibiting a helical or sigmoid shape \citep{2008ApJ...679..827A,2008A&A...481L..65M,2012SoPh..280..475S,2012ApJ...756..124A,2016ApJ...823L..19Z,2017A&A...600A..37N,2019ApJ...874..131A}. The effect of non-planar geometry of coronal loops on oscillation properties was first investigated by \citet{2011A&A...529A..33R}, who considered the loop axis to be a part of a helical line, and of the circular cross-section. They found that the loop non-planarity only weakly affects the estimates obtained through coronal seismology \citep{2012SoPh..278..177S}, and that the simplest model of a straight homogeneous magnetic cylinder provides sufficiently accurate estimates. However, they used an asymptotic method with the ratio of the loop cross-section radius to the loop curvature radius as a small parameter. As stated by the authors, this implies that neither the loop curvature nor the loop twist can directly affect the period of the loop kink oscillations. Instead, the period was affected only indirectly through modifying the dependence of the density on the coordinate along the loop. Moreover, neither these asymptotic results have been tested numerically nor the interference of the effects of non-planarity and the elliptic cross-section has been studied. Previous three-dimensional (3D) numerical simulations of coronal loop oscillations included the effects of curvature \citep{2004ApJ...614.1042M,2006ApJ...650L..91T,2008ApJ...682.1338M,2014ApJ...784..101P}, but maintained a planar loop geometry. In this study, we aim to improve on previous simulations of 3D curved coronal loop oscillations, by including the effect of non-planarity, i.e. considering sigmoid coronal loops. The paper is organized as follows: In Section~\ref{two}, we present the model and the numerical methods; in Section~\ref{three} the simulation results are presented, and in Section~\ref{four} conclusions and implications for coronal seismology are drawn. 

\section{Numerical model and method}
\label{two}

The 3D numerical model consists of a background coronal plasma in a hydrostatic equilibrium in which we embed a coronal loop of higher density. The magnetic field is produced by two magnetic constant-$\alpha$ `poles' of equal intensity and opposite signs located below the simulation region at $x_{0,1}=d$ and $x_{0,2}=-d$, respectively, along the $y=0,z=-z_0$ axis. Here, $\alpha$ is the force-free parameter which controls the helicity of the field lines originating in the poles. The force-free magnetic field of a pole with constant-$\alpha$ is given by \citep{1977ApJ...212..873C, 1989A&A...216..265C}:
\begin{align}
 &B_{x,j} = \frac{1}{r_j^2}\left[ x_j \frac{r_j^2}{R_j^3} \mathrm{cos}(\alpha R_j) - \alpha \frac{x_j z_j^2}{R_j^2} \mathrm{sin} (\alpha R_j) + \alpha x_j \mathrm{sin} (\alpha z_j) + \alpha \frac{y z_j}{R_j} \mathrm{cos}(\alpha R_j) - \alpha y\ \mathrm{cos}(\alpha z_j)\right],\\
 &B_{y,j} = \frac{1}{r_j^2}\left[ y \frac{r_j^2}{R_j^3} \mathrm{cos}(\alpha R_j) - \alpha \frac{y z_j^2}{R_j^2} \mathrm{sin} (\alpha R_j) + \alpha y\ \mathrm{sin} (\alpha z_j) - \alpha \frac{x_j z_j}{R_j} \mathrm{cos}(\alpha R_j) + \alpha x_j\ \mathrm{cos}(\alpha z_j)\right],\\
 &B_{z,j} = \frac{z_j}{R_j^3}[ \mathrm{cos}(\alpha R_j) + \alpha R_j\ \mathrm{sin}(\alpha R_j)],
\end{align}
where $j=1,2$ designates the two poles, $R_j$ is the distance from the position of the $j\mathrm{th}$ magnetic pole to the point at which $\mathbf{B}_j$ is calculated. Furthermore
\begin{align}
&R_j^2 = r_j^2 + z_j^2,\ r_j^2=x_j^2+y^2,\\
&x_j=x-x_{0,j},\ z_j=z-z_0.
\end{align}
The resultant magnetic field from the two opposite sign poles is then:
\begin{align}
&B_x = A (B_{x,1}-B_{x,2}),\\
&B_y = A (B_{y,1}-B_{y,2}),\\
&B_z = A (B_{z,1}-B_{z,2}),
\end{align}
where $A = 1\ \mathrm{kG}$ is the magnetic field intensity.
We use different values for the $\alpha$ parameter. The coronal loop is added by tracing a single magnetic field line, and then using it as a central axis to construct a tube, which is filled with a higher density plasma than the background value. The origin of this single field line, which varies depending on $\alpha$, is chosen in order to maximize the sigmoidity of the resulting loop while keeping it in the simulation domain. The plasma density is gravitationally stratified in both the loop and the background. Note that the background plasma is hotter, therefore the density scale height is higher than inside the loop. Thus, the ratio of densities inside and outside the loop is decreasing with height. 
\begin{figure}[h]
    \centering     
        \includegraphics[width=1.0\textwidth]{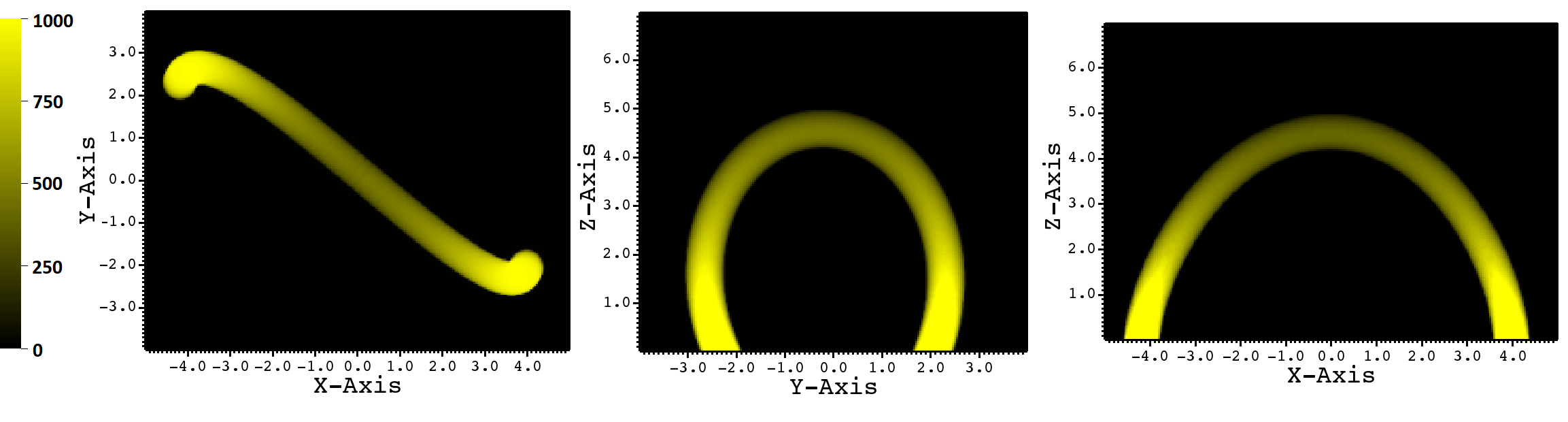}  
        \caption{Synthetic AIA 171\AA$\ $images of the coronal loop for $\alpha = 0.2$, from three different viewpoints corresponding to the three coordinate axes. Values are in user units. }
        \label{synth}
 \end{figure}
Note that the constructed cylindrical tube or loop is not tracing exactly a true flux tube, but it approximates one. In fact, starting from a circular cross-section at one loop footpoint, the cross-section of the flux tube gradually transitions to an elliptical shape towards the apex. In the case of non-zero-$\alpha$, the transition to an elliptical cross-section is asymmetric, and results in an elliptical footpoint at the other end. This effect, which is more pronounced for higher values of $\alpha$, is neglected in the present analysis. Moreover, the flux tube expands with height, which is easily accounted for in the zero-$\alpha$ setup, but it leads to non-trivial expansion for non-zero-$\alpha$ setups, therefore it is also neglected. The implications of this caveat on our results are discussed in Section~\ref{three}.  
\begin{figure}[h]
    \centering     
        \includegraphics[width=1.0\textwidth]{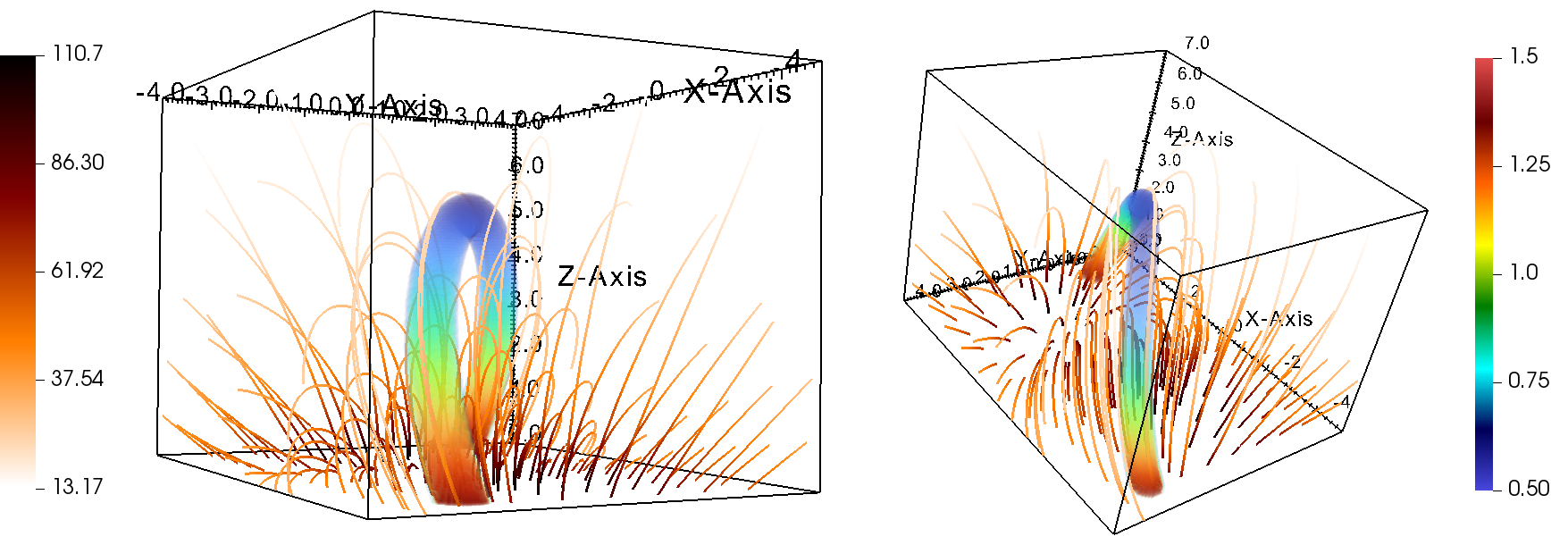}  
        \caption{3D views of the background magnetic field structure and the embedded higher density loop, for $\alpha = 0.2$. The magnetic field is represented through streamlines of the vector field originating at the $z=0$ plane. The color of the streamlines reflects the local magnetic field strength in Gauss. The loop is volume-rendered and the color represents density, in user units.}
        \label{loop}
 \end{figure}
Additionally, the weight of the higher density loop is not counteracted by the force-free magnetic field initially, but the loop starts to fall by a small amount as we start the simulation. This leads to vertically polarized kink oscillations which coexist with the oscillations caused by impulsive horizontal driving at $t=0$. The impulse is a velocity perturbation varying sinusoidally along the loop. The direction of the perturbation is defined by the horizontal perpendicular line to the loop axis in the apex, therefore it is not changing along the loop length. Using this driver, we aim to excite the fundamental standing kink mode of the loop. However, note that the initial perturbation probably does not coincide with the eigenfunction of the fundamental kink, which is not known. For example, in the case of a straight twisted flux tube, it is known that the plane of polarization for the fundamental kink eigenfunction is rotated along the tube \citep{2015A&A...580A..57R}. Therefore, while principally exciting the fundamental kink mode, other modes are also excited to a small degree, including leaky waves. However, in this study, we only investigate the properties of the fundamental standing kink mode. \par 
At the bottom boundary, we set reflective conditions for all three components of the velocity in order to mimic the line-tied conditions of a coronal loop anchored in the dense chromosphere and photosphere. The density and pressure are extrapolated to satisfy the hydrostatic equilibrium in the top and bottom boundaries. All other variables obey a zero-divergence, `continuous' condition. At the lateral boundaries all variables satisfy the `continuous' condition, thus leaving any leaky waves out of the simulation domain. The detailed values of the principal physical parameters are given in Table~\ref{paramtable}
  \begin{table}[h!]
  \caption{The values of principal physical parameters used in the simulations}
  	\centering
  	\begin{tabular}{lccc}
  		\hline\hline
  		Parameter & Value \\ 
  		\hline
  		Box size ($\mathrm{L\times l\times h}$)& $100\times 80 \times 70$ \ Mm \\
  		Pole depth and distance & $z_0 = 30$\ Mm; $d = 30$\ Mm \\
  		Traced field line origin\\
  		(for increasing $\alpha$) $(x,y,z=0)$ & $(35,0); (38,-5); (38,-10); (40,-20)$\ Mm \\
  		Values for the $\alpha$-parameter & $0,0.05,0.1,0.2$ \\
  		Loop radius at footpoint ($R$)& 2.0 Mm \\
  		Loop footpoint density ($\rho_{fi}$) & $3.5 \cdot 10^{-12}$ kg/m$^3$ \\
  		Density ratio at footpoint ($\rho_{fi}/ \rho_{fe}$)& 3 \\
  		Loop temperature & 1.0 MK \\
  		Background plasma temperature & 3.0 MK \\
  	    Average plasma $\beta$ & 0.018 \\
  		\hline
  	\end{tabular}
  	\label{paramtable}
  \end{table}
 \subsection{Method}
The ideal MHD equations are solved in a 3D rectangular domain using \texttt{MPI-AMRVAC 2.0} \citep{2018ApJS..234...30X}, with a finite-volume approach. We applied a splitting strategy for the magnetic field, where the time-independent magnetic field originating from the two `poles' is considered a force-free background field. Thus we only solve for the (nonlinear) perturbed magnetic field components. For this, we used the newly-implemented HLLD solver adapted for magnetic field decomposition described in \citet{2016JCoPh.327..543G}. We use the available second-order 'cada' slope limiter \citep{Cada2009CompactTL}. The base resolution is $48 \times 32 \times 32$. We use four levels of refinement, with the refinement criteria being the density in L\"ohner's error estimator. A refinement study was not completed, however, based on our previous studies of kink oscillations \citep[e.g.,][]{2016A&A...595A..81M}, the period is not affected significantly by resolution. Resolution mostly affects the damping rates, which are not the focus of the present study. The solenoidal condition on the perturbed magnetic field is maintained by combining the implemented parabolic cleaner of Linde and Powell's divergence wave.\footnote{See \url{http://amrvac.org/md_doc_par.html} for more information.} 

\section{Results and Discussion}
\label{three}

Analysis of the oscillation properties is based on synthetic 171\AA \ intensity and Doppler shift images with the lines of sight corresponding to the coordinate axes. Saving synthetic spectroscopic and imaging information about simulations is a built-in feature of \texttt{MPI-AMRVAC}. The magnetic field and density measurements are carried out along the single magnetic field line which is used to construct the denser loop. In order to infer magnetic field estimates seismologically, we need to calculate the theoretical kink speed. For a straight cylinder, this is defined as \citep{1983SoPh...88..179E}:
\begin{equation}
\label{kinkspeed}
c_k = \sqrt{\frac{\rho_i V_{A,i}^2 + \rho_e V_{A,e}^2}{\rho_i + \rho_e}},
\end{equation}
where $V_A$ is the Alfv\'en speed, and the subscripts denote values inside and outside of the coronal loop. Then the period of the fundamental kink mode is given by:
\begin{equation}
\label{kinkperiod}
P = \frac{2 L_\mathrm{loop}}{c_k},
\end{equation}
where $L_\mathrm{loop}$ is the loop length, and the factor two signifies that the wavelength of the fundamental mode is double the loop length.
However, a straightforward application of the above formula is not possible, as the kink speed is varying along the loop. Instead, we calculate the period using the WKB approximation:
\begin{equation}
\label{theorperiod}
P = 2 \int_{s_a}^{s_b} \frac{ds}{c_k(s)},
\end{equation}
where $s_{a,b}$ denote the two footpoints, and $s$ is the coordinate along the loop axis. The measured and theoretically calculated periods are given in Table~\ref{periods}. 
 \begin{table}[h!]
  \caption{The measured period, the length of the centrally traced loop magnetic field line, and the theoretical oscillation period calculated from Eq.~\ref{theorperiod}}
  	\centering
  	\begin{tabular}{c|ccc}
  		\hline\hline
 \text{$\alpha$} & \text{Loop length (Mm)} & \text{Period (s)} & \text{Theor. Period (s)}\\
\hline
 0 & 126.7 & 202.5 & 202.5 \\
 0.05 & 148.3 & 289.3 & 305.4 \\
 0.1 & 150.8 & 251 & 264.7 \\
 0.2 & 148.9 & 120.3 & 145.4 \\
 \hline
 \end{tabular}
 \label{periods}
\end{table}
The measured oscillation periods vary significantly for different $\alpha$ values. This variation is mostly explained by the maximal height of the loops, which is changing with $\alpha$ and the choice of the footpoint. The dipolar magnetic field is rapidly decreasing with heigh, thus lower lying loops have a more intense magnetic field. 
As we can see, in the case of zero $\alpha$, the match between the measured period and calculated period is exact. This demonstrates the robustness of Eq.~\ref{theorperiod}, even when applied to loops which do not satisfy the assumptions used in its calculation. For $\alpha = 0$, loop curvature, loop expansion, and gravitational stratification are the main additional factors. While the loop curvature does not appear to have a significant effect \citep{2009SSRv..149..299V}, the other two factors are known to have a strong effect on the oscillation period. This is understandable, as both factors impact the Alfv\'en speed along the loop. For example, an expansion ratio of two (ratio of the tube radii at the apex and footpoints), as for $\alpha = 0$, would result in an increase of the fundamental mode period by more than a factor of two \citep{2008ApJ...686..694R,2008A&A...486.1015V}. This effect manifests in the decrease of the magnetic field towards the apex, and as such it is accounted for in Eq.~\ref{theorperiod}. In a similar manner, the effect of density stratification \citep{2005ApJ...624L..57A,2005A&A...430.1109A,2005A&A...441..361A} is thus also accounted for. \par
For the simulations with non-zero $\alpha$, we see a general overestimation of the theoretical period which increases with $\alpha$. As mentioned, for non-zero-$\alpha$ we neglect the expansion and ellipticity of the flux tube. This implies that the filling factor (i.e. ratio of the higher density loop cross-section to the flux tube cross section) is decreasing with height, resulting in a lower average density inside the flux tube. Therefore, an overestimation of the period could be explained by the underestimation of the measured Alfv\'en speed, as the higher density plasma is assumed to fill the whole flux tube in its calculation. However, this explanation is in contradiction with the observation that the period overestimation and the filling factor both grow with the value of $\alpha$. While for $\alpha = 0.05$ the filling factor at the apex is 0.35, for $\alpha = 0.2$ the cross section of the flux tube, albeit highly elliptical and with a longer semi-major axis than the loop radius, has a smaller area than the higher density loop. The cross-sectional ellipticity of the flux tube could also explain the overestimation of the period \citep{2003A&A...409..287R,2009A&A...494..295E}. Ellipticity can lead to deviations in the fundamental kink period, proportional to its value, of up to 15\% when compared to a circular cross-section model. However, in our model the ellipticity is varying along the flux tube, making a direct comparison to previous theoretical results difficult. On the other hand, while even for $\alpha = 0$ the flux tube is slightly elliptic at the apex, ellipticity in our model grows with the value of $\alpha$, and therefore can be party considered as a direct result of the sigmoidity of the magnetic field. In this sense, in the present model we attribute the overestimation to the effect of sigmoidity of the magnetic field on the fundamental kink oscillation. \par
\begin{figure}[h]
    \centering     
        \includegraphics[width=0.65\textwidth]{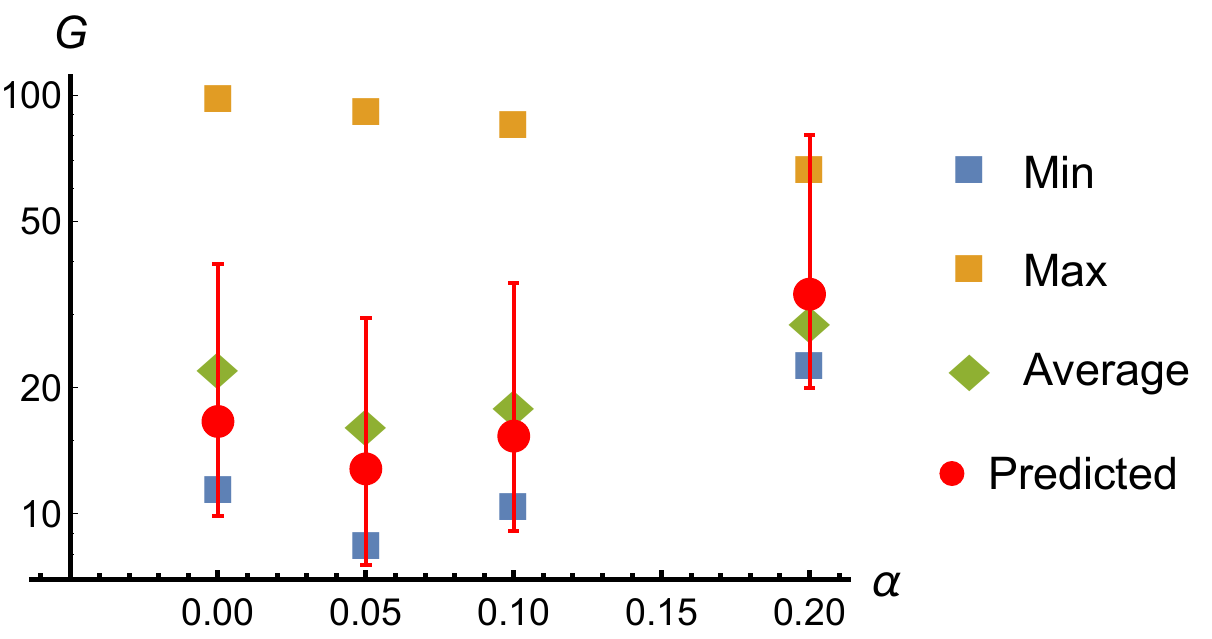}  
        \caption{Plot showing the actual minimum, near the apex (\textit{blue square}), maximum, near the footpoints (\textit{yellow square}), average (\textit{green diamond}), and seismologically estimated (\textit{red circle}) magnitude of the magnetic field inside the oscillating loop, for different values of the $\alpha$ parameter. The seismologically estimated value uses the average value of internal density and density ratio. Values shown are for the single traced field line used to construct the loop. The average is taken over the full loop length. The error bar extends to the highest/lowest estimate resulting from the range of values considered for internal density and the density ratio (details in the text). Note that the ordinate axis is logarithmic.} 
        \label{Bfield}
 \end{figure}

In the following, the implications for coronal seismology will be presented. The magnetic field intensity was determined using the measured oscillation period, loop length, and estimations of the density \citep{2001A&A...372L..53N}:
\begin{equation}
\label{bcalc}
B = \sqrt{\mu_0} \frac{\sqrt{2} L_\mathrm{loop}}{P}\sqrt{\rho_i \left( 1 + \frac{\rho_e}{\rho_i} \right)},
\end{equation}
where $\mu_0$ is the magnetic permeability. The results are shown in Fig.~\ref{Bfield}.
For the estimation of the range of the absolute value of the magnetic field using Eq.~\ref{bcalc} we have considered a range of an order of magnitude for the precision to which the average internal density can be determined. In the current model, the difference between the internal footpoint density and external apex density is less than an order of magnitude. We calculate this range by multiplying the measured average internal density with $10^{\pm \sfrac{1}{2}}$. For the density ratio ($\sfrac{\rho_i}{\rho_e}$) we choose the range $[1.5,10]$. In the simulation, the average density ratio is close to 2, varying along the loop and with the value of $\alpha$. Here we assume that the measurements of the length of the loop and of the period are exact. \par
For $\alpha=0$, despite the similar measured and theoretically calculated periods, the seismologically predicted magnetic field value is lower than the average value. This discrepancy likely comes from the energy density distribution of the fundamental kink mode along the loop \citep{2005A&A...441..361A}, resulting in a weighting function of the form $\mathrm{sin}^2(\pi s/L_{loop})$, which lowers the effective average magnetic field along the loop. In other words, as the displacement amplitude of the fundamental mode has a maximum near the apex, the mode period is more sensitive to the magnetic field near the apex rather than near footpoints. 
For increasing values of $\alpha$, the predicted magnetic field shows an increasing trend relative to the average value, even surpassing it for $\alpha=0.2$. This observation might give rise to the possibility of seismologically determining the sigmoidity of a coronal loop, if some other method to determine the average magnetic field is available, e.g. force-free extrapolations \citep{2012LRSP....9....5W}. Additionally, a relatively accurate determination of the average density inside the loop and of the density ratio would be required. In this sense, the free magnetic energy in a coronal loop could be estimated seismologically. 
The resulting range for seismological estimation of the magnetic field, centred at a value which uses average values of the densities, lies almost entirely within the limits of the measured magnetic field inside the loop. As explained previously, the estimated magnetic field range is shifted towards the minimum value, except for $\alpha = 0.2$, for which case it spans the entire range of magnetic field values measured in the loop.  
  
\section{Conclusion} 
\label{four}
We investigated the effects of coronal loop sigmoidity on the oscillation properties of the fundamental kink mode, and subsequently the impact on coronal seismology. We found that without sigmoidity, the calculated period using WKB theory agrees well with the period found in the simulation. However, increasing the value of $\alpha$ results in theoretical periods increasingly deviating from the measured period. We propose that this dependence could be exploited seismologically in order to measure the non-pontentiality, i.e. the free energy in coronal loops. However, for this method to work, the determination of the average magnetic field along the loop is needed, as well as an accurate measurement of the density along the loop. The average magnetic field could be approximated by, e.g. force-free source surface extrapolation using magnetograms \citep[e.g.,][]{2012LRSP....9....5W}, while the density can be measured using e.g. the emission line ratio method \citep[e.g.][]{1997A&A...327.1230L}. The external/internal density ratio is only weakly impacting the results. On the other hand, we demonstrated the robustness of the seismological method, even when applied to non-planar or sigmoid coronal loops. For all values of sigmoidity considered, the estimation of the magnetic field is within the extremal magnetic field values measured in the loop, despite considering an order of magnitude accuracy for the average density determination.  

\begin{acknowledgements}N.M. was supported by a Newton International Fellowship of the Royal Society. V.M.N. was supported by STFC grant ST/T000252/1. \end{acknowledgements}

\bibliographystyle{apj} % style aa.bst
\bibliography{../Biblio}{} % \begin{tiny}

\end{document}